\documentclass{iopconfser}

\begin{document}

\title{$M1$ Radiative Transition of Light Mesons in the Light-Front Quark Model}

\author{Muhammad Ridwan\orcid{0000-0002-2949-5866}$^{1\dagger}$, Ahmad Jafar Arifi\orcid{0000-0002-9530-8993}$^{2\ddagger}$ and Terry Mart\orcid{0000-0003-4628-2245}$^{1\mathsection}$}

\affil{
$^1$Departemen Fisika, FMIPA, Universitas Indonesia, Depok 16424, Indonesia\\
$^2$Advanced Science Research Center, JAEA, Ibaraki 319-1195, Japan}

\email{$^\dagger$muhammad.ridwan75@sci.ui.ac.id, $^\ddagger$aj.arifi01@gmail.com, $^\mathsection$terry.mart@sci.ui.ac.id}

\begin{abstract}
We investigate the $M1$ radiative transitions between light vector $\mathcal{V}$ and pseudoscalar $\mathcal{P}$ mesons in the 1S and 2S states in the light-front quark model. To this end, we use the light-front wave functions (LFWFs) obtained from the QCD-motivated effective Hamiltonian that includes smeared spin-spin interactions, where a few harmonic oscillator basis functions were employed as trial wave functions.
The results, including their couplings and widths, obtained from LFWFs with different trial wave functions are compared with the available experimental data and other theoretical predictions.
\end{abstract}

\section{Introduction}

Light mesons such as pion ($\pi$), rho ($\rho$), and kaon ($K$), play a central role in understanding the non-perturbative regime of Quantum Chromodynamics (QCD). 
As the lightest quark-antiquark bound states, they provide crucial insights into chiral symmetry breaking, quark confinement, and hadronic structure, serving both as theoretical benchmarks and essential ingredients in the hadronic physics phenomenology. 
To extend our knowledge on phenomenological side of light mesons, we need a phenomenological model to describe the behavior within the meson, i.e., the quark-gluon interaction governed by QCD. 
Additionally, one may scrutinize the meson spectroscopy, since they provide key information about the properties of mesons. 
See Refs.~\cite{Godfrey:1998pd,Godfrey:1985xj} for comprehensive reviews of light mesons.

To understand light mesons, it is important to study their properties, such as mass spectra, decay constants, and transitions,
as they provide insight into the meson's inner structure, especially the light ones. 
In this work, we focus specifically on the $M1$ radiative transition from the light vector ($\rho, K^*$) to pseudoscalar ($\pi, K$) meson or $\mathcal{V} \rightarrow \mathcal{P}\gamma$, as it is a simple yet powerful tool for probing the internal structure via the electromagnetic interaction.
Several approaches have been developed to study radiative transitions in light mesons,  including the relativized quark model~\cite{Godfrey:1985xj}, impulse approximation~\cite{Xu:2024frc}, effective mass scheme~\cite{Mohan:2025kuz}, extended bag model~\cite{Simonis:2016pnh}, relativistic potential model~\cite{Jena:2010zza}, light-front quark model (LFQM)~\cite{Choi:1997iq}, among other phenomenological models. 
In addition, lattice QCD~\cite{Alexandrou:2018jbt} has been used for studying the electromagnetic transition by providing tighter constraints on theoretical models.
Furthermore, several radiative decay channels of ground state light mesons have been observed and listed by the Particle Data Group (PDG)~\cite{ParticleDataGroup:2024cfk}, indicating the need for more reliable phenomenological models to explain and predict for yet-unmeasured channels.

Following the successful computation of the $M1$ radiative transition of heavy quarkonia in the LFQM~\cite{Ridwan:2024ngc,Ridwan:2024hyh} based on light-front dynamics (LFD), we aim to extend our work to light mesons. 
LFD has distinguished features, including a rational energy-momentum relation and seven time-independent kinematic quantities that reflect full Poincar\'{e} symmetries. 
Some efforts have been made based on LFD to investigate the electromagnetic transition in light mesons, including the LFQM~\cite{Jaus:1991cy,Choi:1997iq} and the light-front holographic approach~\cite{Ahmady:2020mht}. 
This work extends previous studies~\cite{Syahbana:2024hkc, Syahbana:2023th}, which focused on the mass spectra and decay constants. The model parameters used in the present analysis are adopted from those studies. 
Different from previous LFQM analyses~\cite{Jaus:1991cy,Choi:1997iq}, we extend them to include the excited states.


\section{Model Description}

In this section, we present the theoretical framework used to describe the meson structure and compute the $M1$ radiative transitions, based on the LFQM.

\subsection{Light-Front Quark Model}

The meson system at rest is described as a bound system of effectively dressed valence quark and antiquark, satisfying the eigenvalue equation of the QCD-motivated effective Hamiltonian, 
\begin{eqnarray} 
H_{q\bar{q}} \ket{\Psi_{q\bar{q}}} = M_{q\bar{q}} \ket{\Psi_{q\bar{q}}},
\end{eqnarray}
where $M_{q\bar{q}}$ and $\Psi_{q\bar{q}}$ are the mass eigenvalue and eigenfunction of the $q\bar{q}$ meson state, respectively. We take the Hamiltonian $H_{q\bar{q}}$ in the quark-antiquark center-of-mass frame as 
$H_{q\bar{q}} = H_0 + V_{q\bar{q}}$ where $H_0 = \sqrt{m_q^2 + \bm{k}^2}  +  \sqrt{m_{\bar{q}}^2 + \bm{k}^2}$ is the kinetic energy part of the quark and antiquark with three-momentum $\bm{k}$. 
The effective inter-quark potential includes a confinement term and two one-gluon exchange terms: the color Coulomb and hyperfine interactions. The obtained eigenfunction will be an input for the LFWF.

The LFWF can be expressed as
\begin{equation} \label{eq:LFWFs}
\Psi^{M}_{q\bar{q}} = \Psi^{JJ_z}_{\lambda_q,\lambda_{\bar{q}}}(x,\bm{k}_{\perp}) = \Phi(x,\bm{k}_{\perp}) \mathcal{R}^{JJ_z}_{\lambda_q,\lambda_{\bar{q}}}(x,\bm{k}_{\perp}),
\end{equation}
where $\Phi(x,\bm{k}_{\perp})$ represents the radial wave function and $\mathcal{R}^{JJ_z}_{\lambda_q,\lambda_{\bar{q}}}(x,\bm{k}_{\perp})$ is the spin-orbit wave function obtained through the Melosh transformation. 
The spin-orbit wave functions $\mathcal{R}^{JJ_z}_{\lambda_q\lambda_{\bar{q}}}$ satisfy the unitary condition, i.e.,
$\Braket{ \mathcal{R}^{JJ_z}_{\lambda_q\lambda_{\bar{q}}} | \mathcal{R}^{JJ_z}_{\lambda_q\lambda_{\bar{q}}} } = 1$.
The LFWF is defined in terms of  Lorentz-invariant internal variables: $x_i = p^+_i / P^+$, $\bm{k}_{\perp i} = \bm{p}_{\perp i} - x_i \bm{p}_{\perp}$, and the light-front helicity $\lambda_i$. 
Here, $P^{\mu} = (P^+, P^-, \bm{p}_{\perp})$ denotes the four-momentum of the meson and $p^{\mu}$ refers to the four-momentum of the $i$-th ($i$=1,2) constituent quark, where we define $x \equiv x_1$ with $\bm{k}_\perp \equiv \bm{k}_{\perp 1}$. 
Then, the three-momentum $\bm{k} = (k_z, \bm{k}_\perp)$ can be written as $\bm{k} = (x, \bm{k}_\perp)$ through the relation
\begin{equation}
k_z = \left( x- \frac12 \right) M_0 + \frac{m^2_{\bar q}-m^2_q}{2M_0}, 
\end{equation}
where the boost-invariant meson mass squared is given by
\begin{eqnarray}
	M_0^2 = \frac{\bm{k}_{\bot}^2 + m_q^2}{x}  + \frac{\bm{k}_{\bot}^2 + m_{\bar{q}}^2}{1-x}.
\end{eqnarray}
Therefore, the variable transformation $\{k_z, \bm{k}_\bot \} \to \{x, \bm{k}_\bot \}$ accompanies the Jacobian factor, 
\begin{equation}
\frac{\partial k_z}{\partial x} = \frac{M_0}{4x(1-x)} \left[ 1 - \frac{ (m_q^2 - m_{\bar{q}}^2)^2}{M_0^4} \right],
\end{equation}
which arises from the change of integration variables.

For radial wave functions, we compare two different scenarios.
First, we employ the two lowest Harmonic Oscillator (HO) basis functions, denoted by $\Phi(x,\bm{k}_{\perp})\equiv \phi_{nS}^{\rm HO}(x,\bm{k}_\bot)$.
The HO basis functions are explicitly given by
\begin{eqnarray} 
	\phi_{1S}^{\rm HO} (\bm{k}) &=& \frac{1}{ \pi^{3/4}\beta^{3/2}} e^{-\bm{k}^2/ 2\beta^2},\\
	\phi_{2S}^{\rm HO} (\bm{k}) &=& \frac{(2\bm{k}^2 -3\beta^2)}{\sqrt{6} \pi^{3/4}\beta^{7/2}} e^{-\bm{k}^2/ 2\beta^2},
\end{eqnarray}
and the corresponding light-front form is given by
\begin{eqnarray}
    \phi_{nS}^{\rm HO}(x,\bm{k}_\bot) = \sqrt{2(2\pi)^3} \sqrt{\frac{\partial k_z}{\partial x}} \phi_{nS}^{\rm HO}(\bm{k}),
\end{eqnarray}
with $\beta$ being a variational parameter inversely proportional to the range of the wave function determined from the mass spectroscopic analysis. 
It should be noted that the wave functions $\phi_{1S}$ include the Jacobian factor $\partial k_z/\partial x$ so that the HO bases $\phi_{nS}$ satisfy the following normalization condition
\begin{eqnarray}
 \int \frac{\dd x~\dd^2 \bm{k}_\bot}{2(2\pi)^3}  |\phi_{nS}^{\rm HO}(x, \bm{k}_\bot)|^2 = 1.
\end{eqnarray}
In the second scenario, we expand the basis as a linear combination of the $1S$ and $2S$ HO bases~\cite{Arifi:2022pal}. 
This can be written in a matrix form as
\begin{eqnarray}
\begin{pmatrix}
\Phi_{1S}  \\
\Phi_{2S} 
\end{pmatrix}
=
\begin{pmatrix}
\cos{\theta} & \sin{\theta}  \\
-\sin{\theta} & \cos{\theta} \\ 
\end{pmatrix}
\begin{pmatrix}
\phi_{1S}^{\rm HO}  \\
\phi_{2S}^{\rm HO} 
\end{pmatrix}
\end{eqnarray}
In principle, a more accurate basis expansion can be carried out. However, in this work, we employ only a few basis functions to investigate the effect of the expansion, given that the HO wave function is commonly used in LFWFs.

\subsection{M1 Radiative Transition}

\begin{figure}[h]
	\centering
	\includegraphics[width=0.6\columnwidth]{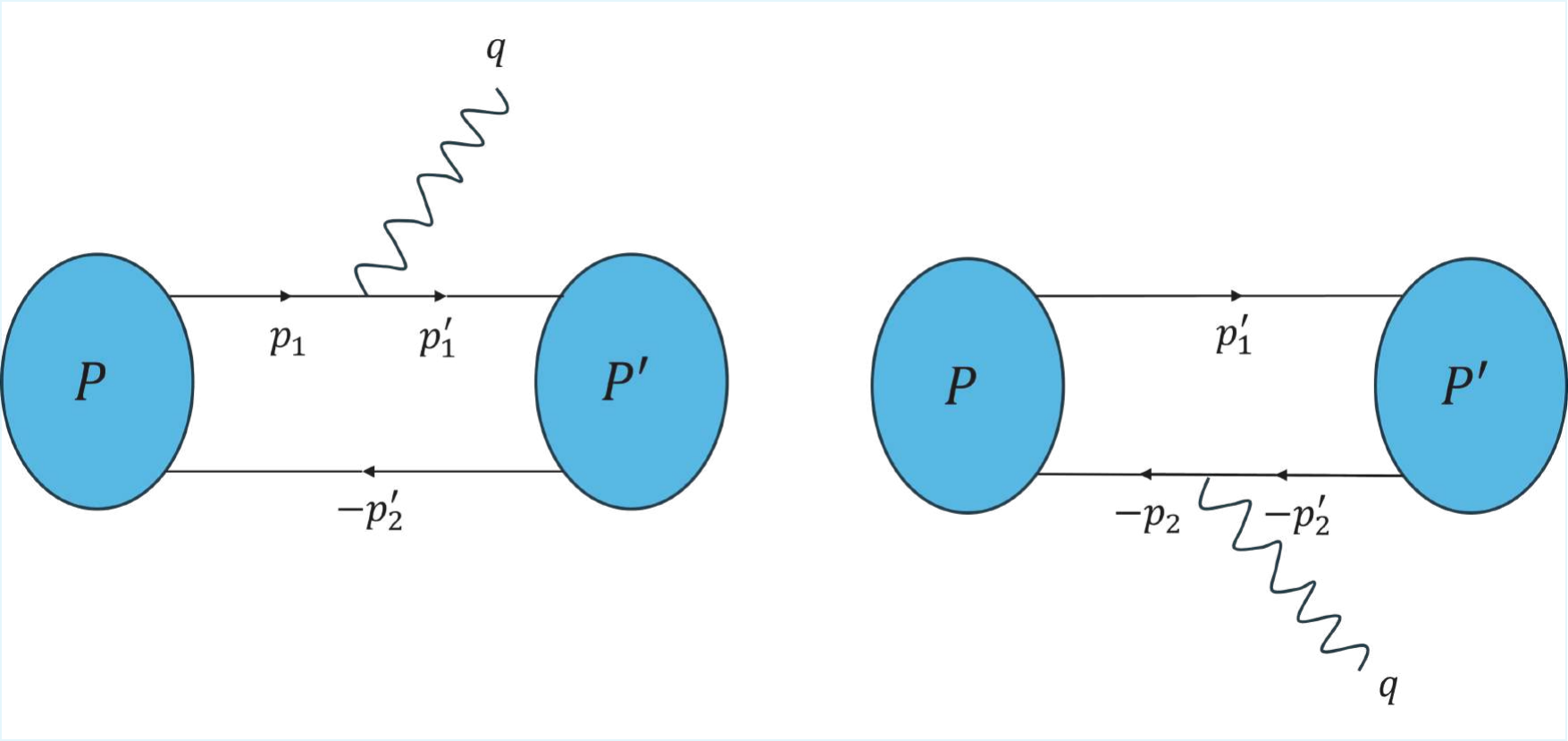} 
	\caption{\label{fig:RD} The Feynman diagrams for the $M1$ radiative transitions $\mathcal{V}(\mathcal{P}) \to \mathcal{P}(\mathcal{V}) \gamma$, where the photon couples to either the quark or antiquark, which are added coherently when computing the decay width.}
\end{figure}

In this work, we focus on this $M1$ radiative transition of light mesons and follow the model description presented in Ref.~\cite{Ridwan:2024ngc}. 
In the LFQM, the Feynman diagram for the $\mathcal{V}(\mathcal{P}) \to \mathcal{P}(\mathcal{V}) \gamma$ process is shown in Fig.~\ref{fig:RD}. 
For the $M1$ transition of the vector meson $\mathcal{V} \to \mathcal{P} \gamma$, the transition form factor  $F_{\mathcal{VP}}(Q^2)$ is defined as
\begin{eqnarray}
\label{eq:rad_M1}
\bra{\mathcal{P}(P^\prime)} J_{\rm em}^\mu(0)\ket{\mathcal{V}(P,h)} = ie \varepsilon^{\mu \nu \rho\sigma} \epsilon_\nu q_\rho P_\sigma F_\mathcal{VP}(Q^2),\quad 
\end{eqnarray}
where the antisymmetric tensor $ \varepsilon^{\mu \nu \rho\sigma} $ assures electromagnetic gauge invariance, and $q = P-P^\prime$ is the four-momentum of the virtual photon. 

The matrix element $\mathcal{J}_h^\mu\equiv\bra{\mathcal{P}(P^\prime)} J_{\text{em}}^{\mu} \ket{\mathcal{V}(P,h)}$ can be expressed in terms of the convolution formula of the initial and final LFWFs as
\begin{eqnarray}
\mathcal{J}_h^\mu = \sum_j e e_q^j \int \frac{\dd x\ \dd^2\bm{k}_\perp}{2(2\pi)^3}  \Phi(x, \bm{k}_\perp^\prime) \Phi(x,\bm{k}_\perp) \sum_{\lambda,\lambda^\prime,\bar{\lambda}} \mathcal{R}_{\lambda^\prime \bar{\lambda}}^{00\dagger}(x, \bm{k}_\perp^\prime) \frac{\bar{u}_{\lambda^\prime}(p_1^\prime) }{\sqrt{x}} \gamma^{\mu} 
\frac{u_{\lambda}(p_1) }{\sqrt{x}} \mathcal{R}_{\lambda \bar{\lambda}}^{1h}(x, \bm{k}_\perp),
\end{eqnarray}
where $e_q^j$ is the electric charge for $j$-th quark flavor $e_c (e_b) = 2/3 (1/3)$.
Note that the invariant mass in the final state becomes
\begin{eqnarray}
    M_0^\prime(x,\bm{k}_\perp^{\prime}) = \frac{m_q^2 + \bm{k}_\perp^{\prime 2}}{x} +  \frac{m_{\bar{q}}^2 + \bm{k}_\perp^{\prime 2}}{1-x}.
\end{eqnarray}
By matching the left- and right-hand sides of Eq.~\eqref{eq:rad_M1},
the form factor can be extracted from
\begin{eqnarray}
    F_\mathcal{VP}(Q^2) = \frac{\mathcal{J}_h^\mu}{\mathcal{G}_h^\mu},
\end{eqnarray}
with $\mathcal{G}_h^\mu=ie \varepsilon^{\mu \nu \rho\sigma} \epsilon_\nu q_\rho P_\sigma$.
The form factor can be separated into two contributions
\begin{eqnarray}
\label{Eq:FF}
F_\mathcal{VP}(Q^2) = e_q I^{\mu}_h(m_q,m_{\bar{q}},Q^2) - e_{\bar{q}} I^{\mu}_h(m_{\bar{q}},m_q,Q^2),
\end{eqnarray}
which represents a process where the photon couples to the quark and antiquark, respectively, as illustrated in Fig.~\ref{fig:RD}.
The one-loop integral $I^{\mu}_{h}$ is given by
\begin{eqnarray}
\label{eq:operator}
 I^{\mu}_{h} = \int \frac{\dd x~\dd^2 \bm{k}_\perp }{2(2\pi)^3} \frac{\Phi(x,\bm{k}_\perp^\prime) \Phi(x,\bm{k}_\perp)}{\sqrt{\mathcal{A}^2+\bm{k}_\perp^{\prime 2}} \sqrt{\mathcal{A}^2+\bm{k}_\perp^2}  } \mathcal{O}_{\mathcal{VP}\gamma}^\mu(h),
\end{eqnarray}
where $\mathcal{A}=x m_{\bar{q}} + (1-x) m_q$ and the operators are given by
\begin{align}
     \mathcal{O}_{\mathcal{VP}\gamma}^+(\pm1) &= 2(1-x)\left(\mathcal{A} + \frac{\bm{k}^2_{\perp}}{\mathcal{D}_0}\right), \\
 \mathcal{O}_{\mathcal{VP}\gamma}^{R(L)}(0) &= \frac{\mathcal{A}}{xM_0}\left(\mathcal{A} +\frac{2\bm{k}^2_{\perp}}{\mathcal{D}_0}\right).  
\end{align}
The partial decay width for the $\mathcal{V} \to \mathcal{P}\gamma$ transition is then computed as
\begin{eqnarray}
\label{eq:DWtheo}
	\Gamma(\mathcal{V}\to \mathcal{P}\gamma) = \frac{\alpha_\mathrm{em}}{(2J_{\mathcal{V}}+1)} g^2_{\mathcal{VP}\gamma} k^3_\gamma,
\end{eqnarray}
where $g_{\mathcal{VP}\gamma}=F_{\mathcal{VP}\gamma}(Q^2=0)$ and $\alpha_\mathrm{em}=1/137$ is the EM fine-structure constant and 
\begin{eqnarray} 
\label{eq:k_photon}
	k_\gamma = \frac{(M_\mathcal{V}^2- M_\mathcal{P}^2)}{2M_\mathcal{V}}
\end{eqnarray}
is the kinematically allowed three-momentum $k_\gamma=\abs{\bm{k}_\gamma}$ of the outgoing real photon.

\section{Result and Discussion}

Before computing the $M1$ radiative transitions, 
we briefly discuss the model parameters adopted from Ref.~\cite{Syahbana:2024hkc},
in which two scenarios ($\theta = 0^\circ$ and $10^\circ$) were considered to analyze the predicted mass spectra, decay constants, and distribution amplitudes of the $\pi$ and $\rho$ mesons using variational analysis.
As shown in Table~\ref{tab:parameters}, we adopt only the parameters obtained with the smeared spin-spin interaction, rather than the contact interaction, as the former provides more reasonable results for the mass spectra and decay constants of excited mesons when compared with experimental data.
Since the $K^*$ and $K$ mesons are also included in our analysis, we use the strange quark mass $m_s$ and the $\beta$ parameter values for the kaon from Ref.~\cite{Syahbana:2023th}.

\begin{figure}[h]
	\centering
	\includegraphics[width=0.95\columnwidth]{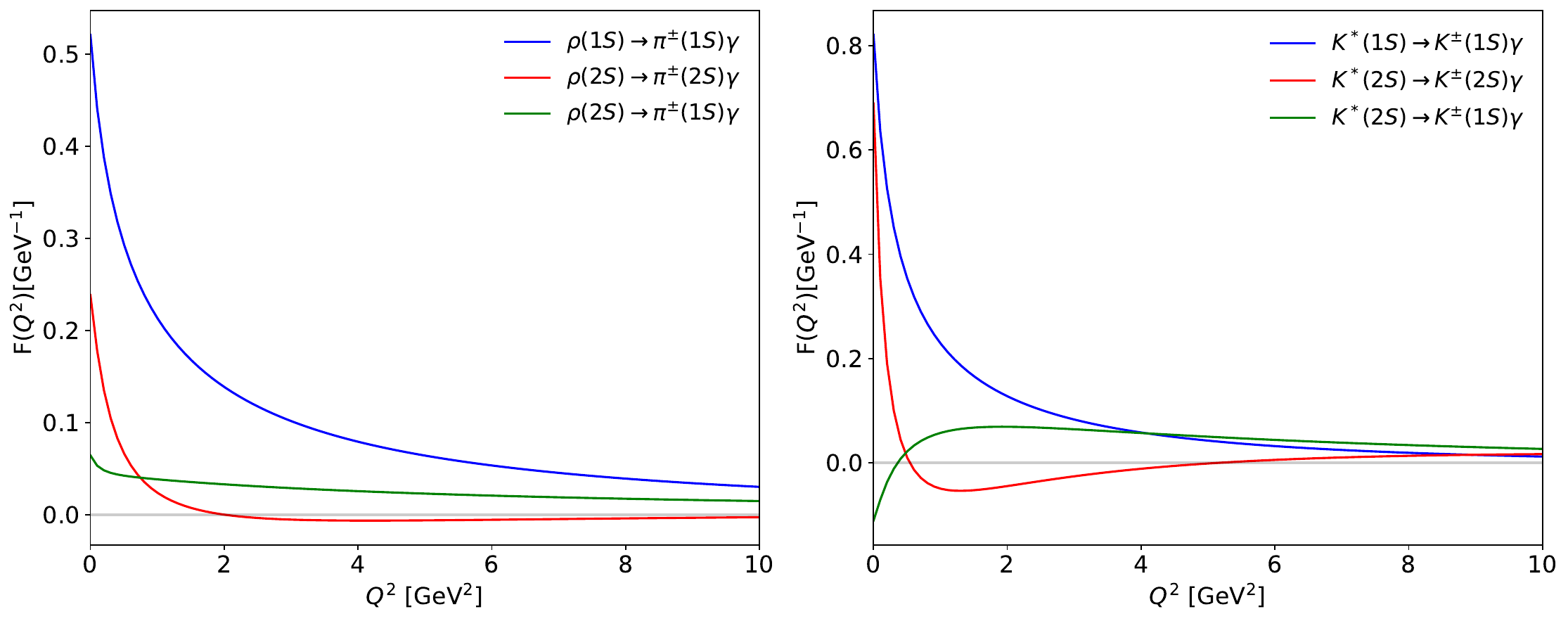}
	\caption{\label{fig:TFF} Transition form factors of light mesons in the $\mathcal{V}(nS) \to \mathcal{P}(n'S) + \gamma$ process, where the blue and red solid lines represent the allowed transitions ($n' = n$) and the green solid line represents the hindered transition ($n' \neq n$). The left panel shows the results for the $\rho$ meson, while the right panel corresponds to the charged kaon. }
\end{figure}


\begin{table}[htb]
	\centering	
	\caption{ Model parameters in the unit of GeV, adopted from Refs.~\cite{Syahbana:2024hkc,Syahbana:2023th}.}
	\label{tab:parameter}
	\begin{tabular}{c|cccccccccc}
		\hline
		$\theta$ & $m_q$ & $m_s$ & $\beta_{\pi}$ & $\beta_{\rho}$ & $\beta_{K}$ & $\beta_{K^*}$ \\ 
		\hline
            $0\degree$  &$0.18$ & $0.34$  &   $0.614$ & $0.318$ & $0.511$ & $0.352$ \\
		$~10\degree~~$ & $~~0.18~~$ & $~~0.34~~$  & $~~0.537~~$ & $~~0.276~~$ & $~~0.466~~$ & $~~0.306~~$ \\
		\hline
	\end{tabular}
	\label{tab:parameters}
\end{table}

To compute the coupling constant $g_{\mathcal{VP}\gamma}$, we set $Q^2 = 0$ in Eq.~\eqref{Eq:FF}, and the results are presented in Table~\ref{tab:results}.
In this calculation, we consider two different current components, namely the plus ($+$) and transverse ($\perp$) currents, and confirm that both components yield the same values for light mesons, consistent with Ref.~\cite{Ridwan:2024ngc}.
We provide the coupling constants for two different scenarios (with different values of $\theta$).
It is evident that increasing $\theta$ generally decreases the value of $g_{\mathcal{VP}\gamma}$, except for the transition $\rho(2S) \rightarrow \pi(2S)\gamma$, where the value increases moderately.
Note that we also include the neutral kaon for the ground state in the coupling and width because the availibility of experimental data as well as other theoretical models.
A similar trend is observed in $K(2S) \rightarrow K(2S) \gamma$ and $K(2S) \rightarrow K(1S) \gamma$, although in the latter case, the coupling constant $g_{K^*(2S) \rightarrow K(1S)\gamma}$ is negative, indicating a destructive interaction in this hindered kaon transition.
Compared to the extended bag model (EBM)~\cite{Simonis:2016pnh} and the relativistic potential model (RPM)~\cite{Jena:2010zza}, our results are moderately lower in the $\rho(1S) \rightarrow \pi^{\pm}(1S) \gamma$, but are in the good agreement in the $K^*(1S) \rightarrow K(1S)\gamma$, either in neutral or charge ones. 
This can be attributed not only to differences in model parameters but also to the wave functions employed: the EBM and RPM use basis functions derived from the free Dirac equation, where the radial wave functions are characterized by Bessel functions, particularly in the $S$ wave.
Interestingly, the value of $g_{\rho(1S) \rightarrow \pi(1S), \gamma}$ extracted from lattice QCD~\cite{Alexandrou:2018jbt} is $0.607$ GeV$^{-1}$, obtained by dividing the lattice result by the pion mass $m_\pi = 0.135$ GeV to match our definition. Overall, our results for ground state kaon show better agreement with other theoretical models than those for rho mesons. 

Figure~\ref{fig:TFF} shows the transition form factor for the $\rho$ and $K$ mesons. It is evident that the $\rho^{\pm}(2S) \rightarrow \pi^{\pm}(2S) \gamma$ transition (red solid line) drops more steeply than the $\rho^{\pm}(2S) \rightarrow \pi^{\pm}(1S) \gamma$ transition (green solid line), tending toward negative values as $Q^2$ increases. Compared to the those for heavy quarkonia, $g_{\rho(2S) \rightarrow \pi(2S) \gamma}$ approaches zero at $Q^2 = 2$ GeV$^2$, whereas $g_{\psi(2S) \rightarrow \eta_c(2S) \gamma}$ (shown in Fig.~7 of Ref.~\cite{Ridwan:2024ngc}) vanishes at around $Q^2 = 5$ GeV$^2$. 
The differences in the form factors between light and heavy mesons arise mainly from the quark masses and the $\beta$ parameters, which characterize the size of the wave function. Moreover, in allowed transitions, the initial and final states overlap, while in hindered transitions, the two states are orthogonal. Because of this, the hindered transition $\rho^\pm \to \pi^\pm \gamma$ yields a coupling constant with a positive sign, unlike the kaon and heavy mesons, which produce a negative sign, as shown in Ref.~\cite{Ridwan:2024ngc}. 
This behavior is partly due to the large difference in the value of $\beta$ parameters for $\rho$ and $\pi$ mesons.
This demonstrates the sensitivity of the hindered transition to the meson LFWFs.

\begin{table*}[htb]
	\centering
 		\renewcommand{\arraystretch}{1.3}
	\caption{Coupling constants $g_{\mathcal{VP}\gamma}$ [GeV$^{-1}$] and partial decay width $\Gamma$ [KeV] in the present LFQM compared with results from other models: EBM~\cite{Simonis:2016pnh}, and RPM~\cite{Jena:2010zza}. } 
	\label{tab:results}
	\begin{tabular}{c|c c c c|c c c c}
		\hline\hline
		Transition & $g_{\theta = 0}$ & $g_{\theta = 10\degree}$ & \cite{Simonis:2016pnh} & \cite{Jena:2010zza} &$\Gamma_{\theta = 0}$ & $\Gamma_{\theta = 10\degree}$ & \cite{Simonis:2016pnh} & \cite{Jena:2010zza} \\
		\hline\hline
		$\rho^{\pm}(1S) \rightarrow \pi^{\pm}(1S) \gamma$  & 0.540  & 0.522 & 0.720 & 0.694 & 36.85 & 34.44 & 76 & 68.12 \\
		$\rho^{\pm}(2S) \rightarrow \pi^{\pm}(2S) \gamma$  & 0.159  & 0.239 & \dots & \dots & 0.233 & 0.525 & \dots & \dots\\
	   $\rho^{\pm}(2S) \rightarrow \pi^{\pm}(1S)\gamma$   & 0.095 & 0.065 & \dots & \dots &  8.430 & 3.878 & \dots & \dots \\
        $K^*(1S) \rightarrow K^{\pm}(1S)\gamma$ & 0.841  & 0.823 & 0.905 & 0.882 & 49.8 & 47.7 & 68 & 63.2 \\ 
        $K^*(1S) \rightarrow K^0(1S)\gamma$ & $-1.198$ & $-1.177$ & $-1.20$ & 1.23 & 101 & 97.4 & 134 & 123.4 \\
        $K^{\pm}(2S) \rightarrow K^*(2S)\gamma$ & 0.504  & 0.693 & \dots & \dots & 0.172 & 0.326 & \dots & \dots\\
        $K^*(2S) \rightarrow K^{\pm}(1S)\gamma$ & $-0.087$  & $-0.111$ & \dots & \dots & 4.40 & 7.082 & \dots & \dots \\
		\hline\hline
	\end{tabular}
 		\renewcommand{\arraystretch}{1}
\end{table*}

After obtaining $g_{\mathcal{VP}\gamma}$, we also calculate the corresponding partial decay widths, which are summarized in Table~\ref{tab:results}. We find that the decay width for the $M1$ radiative transition between the ground states is suppressed when $\theta = 10^\circ$ is applied. In contrast, for the ${\rho(2S) \rightarrow \pi(1S)\gamma}$ and ${K(2S) \rightarrow K^(2S)\gamma}$ transitions, the decay widths increase when $\theta = 10^\circ$ is used. Additionally, we observe that $\Gamma_{\rho(1S) \rightarrow \pi(1S) \gamma}$ and $\Gamma_{K^(1S) \rightarrow K(1S) \gamma}$ are significantly larger than those of the other transitions, mainly due to the substantial mass difference between the vector and pseudoscalar mesons, which enhances the photon phase space factor $k_{\gamma}$ in Eq.~(\ref{eq:k_photon}).
According to the PDG, the extracted decay widths for the $\rho$ and $K^*$ ground-state transitions are 69.28 keV and 50.37 keV (neutral channel), respectively, 
while the charged channel has a decay width of 116 keV for $K^*(1S) \rightarrow K^{\pm}(1S) \gamma$. 
In general, these experimental values are slightly larger than our predicted results.
A similar trend is observed in results from the EBM and the RPM, likely due to differences in the model parameters. Likewise, the decay width extracted from lattice QCD~\cite{Alexandrou:2018jbt} for the $\rho(1S) \rightarrow \pi(1S), \gamma$ transition is 84.2 keV, also exceeding our result.

When comparing with another LFQM study with a single Gaussian Ansatz~\cite{Choi:1997iq}, where both HO and linear potential models were used, they obtained 76 keV (HO) and 69 keV (linear) for the $\rho \rightarrow \pi\gamma$ transition, and 79.5 keV (HO) and 71.4 keV (linear) for the $K^* \rightarrow K\gamma$ transition—again, all higher than our predictions.
The reason our predicted values are smaller than both experimental data and other theoretical models is that the model parameters we adopted from Ref.~\cite{Syahbana:2024hkc} are smaller in magnitude, which leads to greater discrepancies in certain transitions.

Overall, we observe that some physical quantities in our results differ from those predicted by other theoretical models. 
This discrepancy may arise from differences in model assumptions, parameter choices, and treatment of interactions.
In the case of light mesons such as the $\rho$ and $K^*$, strong decays are kinematically allowed even in the ground state and are the dominant decay modes (e.g., $\rho \to \pi \pi$, $K^* \to K \pi$).
Nevertheless, electromagnetic processes such as $M1$ transitions still play an important role, especially in probing internal structure and wave function characteristics, where strong decay channels may not directly provide such information.
In contrast, for heavy mesons such as quarkonia, as discussed in Ref.~\cite{Ridwan:2024ngc}, strong decays are often suppressed in the ground state due to being below the open-flavor threshold. 
As a result, electromagnetic decays, such as $J/\psi \rightarrow e^- e^+$ or $M1$ radiative transitions, become the primary decay modes and serve as useful observables for theoretical comparison.

\section{Summary and Outlook}

We have computed the $M1$ radiative transitions of light mesons ($\rho$ and $\pi$) from the ground state up to the first radially excited state using the light-front quark model.
To probe the magnetic dipole transitions, we calculated the coupling constants and partial decay widths for each channel using model parameters adopted from previous studies~\cite{Syahbana:2024hkc,Syahbana:2023th}. 
Our results show relatively large decay widths for $\Gamma_{\rho(1S) \rightarrow \pi(1S) \gamma}$ and $\Gamma_{K^(1S) \rightarrow K(1S) \gamma}$, primarily due to the substantial mass difference between the vector and pseudoscalar mesons involved. We also find that the predicted coupling constants $g_{\rho(2S) \rightarrow \pi(2S) \gamma}$ and $g_{K^(2S) \to K(2S)\gamma}$ increase significantly when the mixing angle is set to $\theta = 10^\circ$, although this enhancement is suppressed as $Q^2$ increases. In the hindered transitions, we observe a negative coupling for the kaon meson, while the $\rho$ meson yields a positive value.

For future work, we plan to expand more basis functions to obtain more realistic LFWFs and explore potential differences in magnetic dipole transitions of light mesons. 
Additionally, we aim to re-fit the model parameters using up-to-date experimental data, as our current parameter set yields predictions that are relatively smaller than those of other models.

\section*{Acknowledgments}

M.R. thanks the IPS 2025 organizer for providing the oral presentation for his work. This work has been supported in part by the PUTI Q1 grant from the University of Indonesia under contract No. NKB-441/UN2.RST/HKP.05.00/2024.

\end{document}